\documentclass[floatfix,prd,lengthcheck,showpacs,amssymb,amsmath,amsfonts,aps,altaffilletter,nofootinbib,nopreprintnumbers,showpacsm]{revtex4-1}

\usepackage[utf8]{inputenc}
\usepackage[dvipsnames]{xcolor}
\usepackage{graphicx}
\usepackage{subfigure}
\usepackage{amsmath}
\usepackage{acronym}
\usepackage{multirow}
\usepackage{tabu}
\usepackage{import}
\usepackage{hyperref}
\hypersetup{
    colorlinks=true,
    linkcolor=blue,
    filecolor=magenta,      
    urlcolor=cyan,
}

\begin{document}

\title[]{Extending the PyCBC search for gravitational waves from compact binary mergers to a global network}

\author{Gareth S. Davies$^1$}
\author{Thomas Dent$^1$}
\author{M\'arton T\'apai$^2$}
\author{Ian Harry$^3$}
\author{Connor McIsaac$^3$}
\author{Alexander H. Nitz$^{4,5}$}

\address{$^1$ Instituto Galego de F\'{i}sica de Altas Enerx\'{i}as, Universidade de Santiago de Compostela, 15782 Santiago de Compostela, Galicia, Spain}
\address{$^2$ Department of Experimental Physics, University of Szeged,
Szeged, 6720 D\'{o}m t\'{e}r  9., Hungary}
\address{$^3$ University of Portsmouth, Portsmouth, PO1 3FX, United Kingdom}
\address{$^4$ Max-Planck-Institut f{\"u}r Gravitationsphysik,
         Albert-Einstein-Institut, D-30167 Hannover, Germany}
\address{$^5$ Leibniz Universit{\"a}t Hannover, D-30167 Hannover, Germany}


\date{\today}

\begin{abstract}
The worldwide advanced gravitational-wave (GW) detector network has so far primarily consisted of the two Advanced LIGO observatories at Hanford and Livingston, with Advanced Virgo joining the 2016-7 O2 observation run at a relatively late stage. 
However Virgo has been observing alongside the LIGO detectors since the start of the O3 run; 
in the near future, the KAGRA detector will join the global network and a further LIGO detector in India is under construction. Gravitational-wave search methods would therefore benefit from the ability to analyse data from an arbitrary network of detectors. In this paper we extend the PyCBC offline compact binary coalescence (CBC) search analysis to three or more detectors, and describe resulting updates to the coincident search and event ranking statistic. For a three-detector network, our improved multi-detector search finds 23\% more simulated signals at fixed false alarm rate in idealized colored Gaussian noise, and up to 40\% more in real data, compared to the two-detector analysis previously used during O2.
\end{abstract}

\maketitle

\section{Introduction}
\label{s:intro}
The recent advent of gravitational-wave astronomy with observations by the Advanced LIGO-Virgo network~\cite{AdvLIGO2015,Acernese_2014} of the coalescence of binary black hole and binary neutron star systems~\cite{PhysRevLett.116.061102,TheLIGOScientific:2017qsa,Abbott_2019} has been made possible by the effective operation of search algorithms to identify scarce signals among months of data dominated by noise.  Although sufficiently high amplitude signals may be detectable by generic (unmodeled) searches for excess power~\cite{TheLIGOScientific:2016uux,Klimenko_2016}, the majority of binary merger signals are only identifiable by employing accurate waveform models in matched filter searches~\cite{Finn:1992xs,PhysRevD.85.122006}, as in recent searches of public LIGO-Virgo data (e.g.\ \cite{Magee_2019}) where low-amplitude events are identified~\cite{PhysRevD.100.023011,2-ogc,Venumadhav:2019lyq}.

The most sensitive searches targeting CBC sources~\cite{PhysRevD.87.024033,Cannon_2012,PhysRevD.95.042001,sachdev2019gstlal,Buskulic_2010,adams2015low,Hooper_2012,Usman_2016,Nitz_2017} detect signals by correlating a bank of waveform templates with the data from each detector in a network, recording peaks in the matched filter signal-to-noise ratio (SNR) time series as \textit{triggers}.  These triggers are then compared to those from other detectors to check for \textit{coincidences} formed with consistent times of arrival, component masses and spins. The resulting coincident events are ranked according to the events' parameters, and then compared to an estimate of the background noise distribution in order to measure the significance of candidate coincident events.

A common challenge in the implementation of templated searches is the high computational cost of correlating months of data at kHz sample rates against $10^5$ -- $10^6$ templates of duration up to minutes.  The PyCBC search algorithm~\cite{Canton:2014ena,Usman_2016}, built on a highly modular and configurable set of libraries~\cite{alex_nitz_2020_3874649}, achieves high efficiency by using Fast Fourier Transform implementations optimized for different computing platforms, and is thus able to identify candidate events with latencies of tens of seconds at moderate computational cost~\cite{Nitz_2018}. PyCBC workflows have been designed to take advantage of diverse computational resources including local clusters, XSEDE, and the Open Science Grid~\cite{Weitzel:2017ocs} using the Pegasus workflow management system~\cite{deelman-fgcs-2015}. 

Until now, the offline (archival) PyCBC coincident search has analyzed data from the two Advanced LIGO detectors only. In general, with data from more than two detectors available within the framework of the triggered coincident search, sensitivity will be optimized by generating and combining triggers from all detectors and accounting for various effects, including which detectors are operating and the detectors' differing sensitivities and antenna patterns. Additionally, the ranking procedure for candidates must account for these effects to preserve search sensitivity.

During part of the initial GW detector era (2002-2010), data from the Virgo detector with 3\,km arm length was analyzed in addition to the two LIGO detectors with 4\,km arm length, and during some of the science runs a co-located 2\,km detector at LIGO Hanford, using the \texttt{ihope}~\cite{PhysRevD.87.024033} analysis, a predecessor of the PyCBC search pipeline architecture.  The resulting four-detector search~\cite{Abadie_2010} was, though, complex and severely limited by computational cost in practice. 
Since then the extension of coincident searches to three or more detectors has been addressed by accounting for signal time of flight between detectors in the MBTA pipeline~\cite{adams2015low} and in the GstLAL pipeline~\cite{gstlal_3det,sachdev2019gstlal} using a ranking statistic evaluated via nearest-neighbor approximation over the parameter space of multi-detector events.

The PyCBC Live (online) search deployed during the O2 Advanced LIGO-Virgo observing run extends the existing two-detector search, following up selected significant two-detector coincidences by calculating the corresponding matched filter time series in any additional detectors and incorporating
this information in candidate significance. However, this procedure is not yet optimized for sensitivity at high thresholds of significance. (Note that all available detectors are also used for source localisation and parameter estimation~\cite{Singer_2016,Biwer_2019}.)

In this paper, we present changes to the PyCBC offline analysis to search data from three or more detectors, to allow all available detectors to generate coincident events and to compare and calculate significance for the resulting different combinations of these triggers.
By using three or more detectors within the coincident search we increase the search duty cycle, as we can form coincidences in any time where at least two detectors are observing. The method is applicable to arbitrarily large networks, however this paper uses the three-detector (LIGO Hanford, LIGO Livingston and Virgo) network for explanations and examples due to the current availability of archived data.

\begin{figure*}[htp]
    \centering
    \includegraphics[width=1.9\columnwidth]{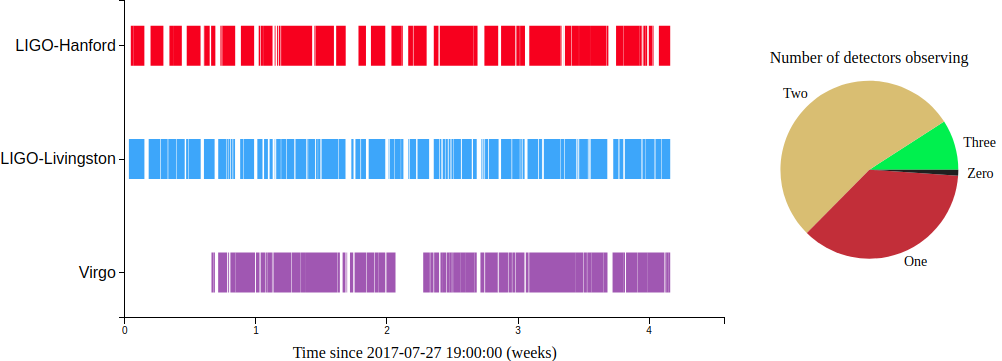}
    \caption{Left: Times for which the detector data was marked as `science ready' during late O2. With Virgo joining during this period, there is a total of 15 days of data where three detectors are observing simultaneously.  Right: Pie chart showing the proportion of time over all of O2 (Nov 30 2016 -- Aug 25th 2017) where a given number of detectors are observing.
    }
    \label{fig:ScienceTime}
\end{figure*}

Figure~\ref{fig:ScienceTime} shows when the detectors were observing during the last few weeks of the joint O2 LIGO-Virgo analysis, including the first period of three-detector observation in the Advanced detector era. Although there is a significant fraction of single-detector time during this period, searches for gravitational wave signals using a single observatory are not considered here; see \cite{Nitz2020search, sachdev2019gstlal, Callister_2017}.

We also improve sensitivity over the two detector search in times where more than two detectors are operating, as it may be possible to obtain coincidences even in cases where for one detector the line of sight to the source lies in a blind spot or the data is temporarily of poor quality. Sensitivity for such times of multi-detector operation is also increased by the ranking of events where triggers are present in three or more detectors, taking advantage of the very low rates of noise coincidences for such detector combinations.

The scheme for finding coincidences between triggers is discussed in section~\ref{s:coinc}. The calculation of significance requires a \textit{ranking statistic}, a function of the matched filter SNRs and $\chi^2$ signal-based veto values in different detectors, and of the intrinsic (mass, spin) and extrinsic (time of arrival, amplitude, phase) properties of the apparent signal, in comparison with the estimated noise background distribution of similar templates. We will discuss the ranking statistic and its development for the case of more than two detectors in section~\ref{s:rankingstat}. In section~\ref{s:significance} we discuss how coincidences in different detector combinations can be compared for obtaining overall significance. Section~\ref{s:senscomp} then shows how these changes to the analysis combine to improve the sensitivity of the network to compact binary coalescence gravitational-wave signals.

\section{Multi-detector coincidence}
\label{s:coinc}
Triggers produced in multiple detectors from a common astrophysical source will occur within a short time window of each other, given by time-of-flight considerations and timing measurement uncertainty. This fact allows us to exclude the vast majority of noise triggers, which are uncorrelated in time between detectors, thus forming the basis of our test for coincidence. A signal would also produce triggers in the same waveform template in all detectors, so we only search for coincidence between detectors in a given template~\cite{Usman_2016}.

We check for these coincidences in all combinations of detectors in the first stages of the analysis. For notation, we will refer to coincidences by the initials of the detectors involved, for example `HL' coincidences are formed by the LIGO Hanford (H) and LIGO Livingston (L) detectors, and an `HLV' coincidence will also incorporate a trigger from Virgo (V). Thus for the LIGO-Virgo network we form coincidences in the combinations HL, HV, LV and HLV.

A coincidence is formed if two or more triggers from different detectors are within a certain time window. This window is taken as the time of flight for gravitational waves between the sites plus a small, fixed amount to allow for timing errors. Two-detector coincidences are found by comparing the times of triggers in the two detectors; if the difference between the arrival times is less than the allowed time window, then they are considered coincident.

\begin{figure}[htp]
    \centering
    \includegraphics[width=0.4\textwidth]{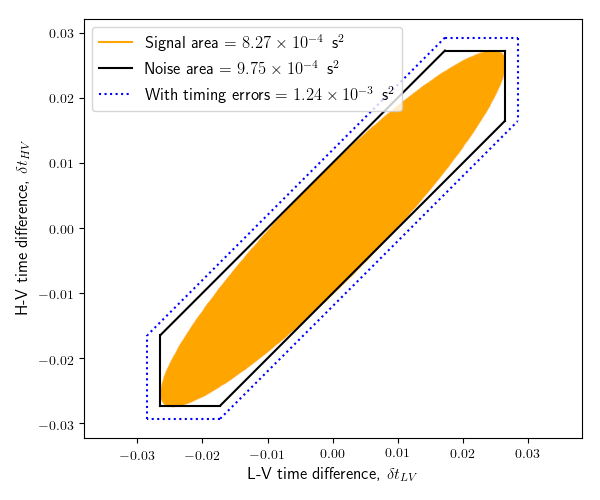}
    \caption{Allowed time differences between LIGO-Livingston and Virgo given the time difference between LIGO-Hanford and Virgo. The black lines show the region where three-detector coincidences are allowed to form, neglecting timing error; dotted blue lines show the allowed area for noise coincidences allowing for 2\,ms timing error. The orange shaded area shows where signals are expected to lie, neglecting timing error. 
    Although coincident events are allowed over a larger time window than is physically possible for signals, events far outside the signal area are suppressed by the phase-time-amplitude consistency checks of section~\ref{sss:PTA}.}
    \label{fig:allowedtimes}
\end{figure}

We form three-detector coincidences by applying the same two-detector coincidence test to trigger time differences for each pair of detectors. Figure~\ref{fig:allowedtimes} shows the allowed differences in arrival times of noise and signal coincidences in three detectors. We see that the allowed window is slightly larger than the one in which signals will fall. The consequences of this multi-detector coincidence test are discussed in more detail in section~\ref{ss:rankstatnoise}. Coincidences in a region not populated by signals but allowed for noise coincidences will be heavily down-ranked through the time difference consistency part of the phase-time-amplitude consistency checks of section~\ref{sss:PTA}.

\subsection{Multi-detector background distribution estimation}
\label{ss:timeshifts}
In order to measure whether a candidate coincidence is significant we assign it a ranking statistic based on its intrinsic and extrinsic parameters, as detailed below, which we compare to the ranking statistics of a manufactured set of noise coincidences which form the background distribution estimate. 
By counting how many of these manufactured background events are ranked higher than each candidate event and dividing by the effective length of time for which background events were generated, we may calculate false alarm rates (FARs) down to one per tens of millennia for week-long stretches of data.

This manufactured set of noise coincidences is made up of combinations of single detector triggers which have been time-shifted such that the difference between arrival times at different detectors is not physically allowed. These \textit{time-shifted coincidences} are treated in the same way as candidate coincidences in order to estimate the distribution of the ranking statistic under a no-signal hypothesis~\cite{Usman_2016, PhysRevD.87.024033}.

In the two-detector configuration, applying time shifts is straightforward as shifting one detector is entirely equivalent to shifting the other detector in the other direction. Typically we perform many time-shifted background analyses with a regularly spaced set of time-shifts at intervals of $0.1$\,s for HL coincidences, which is a few times larger than the maximum physical time difference.
However in the many-detector configuration, there is the possibility of allowing data from every detector to shift relative to the others. For just a few days of coincident data, given a 0.1\,s time shift interval, applying all possible relative shifts between three detectors would result in time-shifted background analyses producing an amount of coincidences equivalent to the detector network running for the age of the universe, and the resulting data storage would become unreasonable.

\begin{figure}[htp]
    \centering
    \includegraphics[width=\columnwidth]{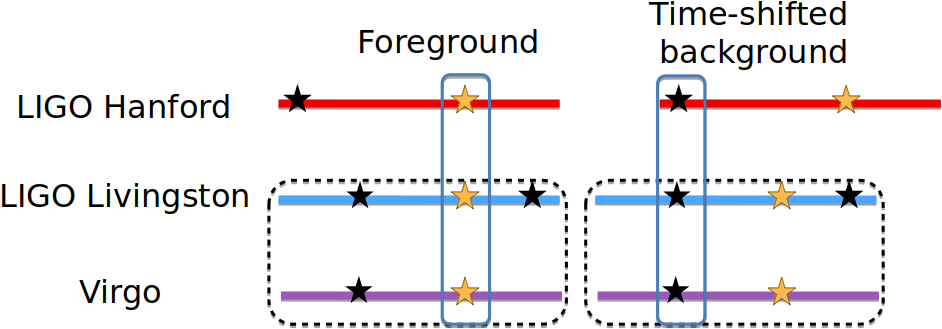}
    \caption{Diagram showing how background coincidences are formed by time shifts for comparison with three-detector coincidences. The shifted detector here is the LIGO Hanford detector, and LIGO Livingston and Virgo are fixed to one another as shown by the dotted rectangle. The stars show triggers in each detector, with gold stars showing triggers which form candidate coincidences and black stars showing triggers which do not. On the left, a candidate coincidence is highlighted by a blue rectangle; on the right, the time-shift procedure is illustrated, with a time-shifted coincidence highlighted. In this procedure the timestamps of Hanford detector triggers are shifted by a fixed offset, allowing a three-detector coincidence to form. This is only possible where a two-detector coincidence is formed within the fixed subset. 
    }
    \label{fig:timeshift_example}
\end{figure}

Thus, we choose to fix all the detectors relative to one another, except for one, whose data is time-shifted relative to all the others. As a consequence, any triple-detector background coincidences will require two-detector foreground coincidences within this fixed set. Figure~\ref{fig:timeshift_example} demonstrates how the time-shifted coincidences are formed.

Other possible configurations would be to perform time shifts of randomized size on all detectors, up to a fixed total number of time-shifted coincidences, to use different time shift increments for different detectors, or to shift the detectors in the fixed set by a constant amount which is different for each detector. These choices are expected to yield similar estimates, and our chosen configuration is the most straightforward. However, with this configuration (and for the time-shift method in general) care is required in order to minimize contamination of triggers used in the time-shifted analyses by loud signals, as we discuss in the next section.

\subsection{Removal of signals from background estimates}
As discussed above, a candidate's significance is measured by its FAR, the expected rate of coincident noise events with a higher ranking statistic. Accurate calculation of significance requires, among other issues, separating loud signal triggers from noise triggers. Time-shifted coincidences which contain a trigger from a known signal do not accurately represent the noise distribution and therefore may bias our significance estimation.

To counter contamination of signals within our FAR calculation we remove triggers from apparent signals from the time-shifted analyses as much as possible. We do this by successively taking the highest ranked candidate coincidence and, if its estimated FAR is below a threshold at which we consider it a confident detection, subsequently removing all triggers from any detector and template, with times within $\pm 1\,$s of the candidate, from the background estimate for any other, lower-ranked candidates~\cite{TheLIGOScientific:2016pea}.
The FAR calculation for a given candidate is thus inclusive of its own triggers and accounts for the hypothesis that the candidate (as well as all lower ranked events) are caused by noise. This removal procedure is repeated until the FAR for the highest ranked remaining candidate event is no longer below the confident detection threshold. A more comprehensive discussion of signal trigger removal from time-shifted analyses is given in~\cite{hamlet}.

For our chosen method of applying time shifts to triple coincident events, we require the triggers from the fixed (non-shifted) detectors to be coincident without applying any time shift. Thus, they also form two-detector candidate events. Therefore a small number of such triggers could arise from signals seen in the fixed sub-network, but not in the shifted detector, which would then contaminate the background estimate for the three-detector combination. This is mitigated in different ways, depending upon whether the shifted detector is observing at the time of such two-detector events, but are relevant to any detector which does not trigger at the time of the candidate coincidence.

To mitigate contamination from signals where a detector was operating, but did not contribute to the candidate coincidence, we do not use triggers from any detectors around the time of any significant candidate coincidences in the background estimate. For example, if a highly-significant candidate is seen in HL, and Virgo is available but did not produce a trigger, then the Hanford and Livingston triggers will also be removed from the HLV, HV and LV background estimates.

To reduce contamination from signals occurring when one or more detectors are not observing, we additionally require that the background estimate for a given detector combination should include triggers only from times when all the detectors involved are observing.
For example, we exclude all triggers which occur when Hanford is not observing from the background estimates of HLV, HL and HV coincidences, regardless of their significance; we also exclude all triggers in single-detector time from any coincident background estimates.

\subsubsection{Choice of shifted detector for background coincidences}
\label{sssec:pivot}

Within our scheme, the choice of which detector to time-shift can strongly influence the amount of contamination by signal triggers in the time-shifted background.  We minimize this contamination by ensuring that the least sensitive detector is within the fixed sub-network: then, it is unlikely that a signal will generate a candidate event in the fixed sub-network without also producing a coincident trigger in the time-shifted detector.

In order to show the effect of the choice of detector being shifted we compare background coincidences for each choice of shifted detector given the same set of single detector triggers. For this demonstration, we use the same six days of data as is used in section~\ref{ss:a20comp}.
\begin{figure}[tb]
    \centering
    \includegraphics[width=\columnwidth]{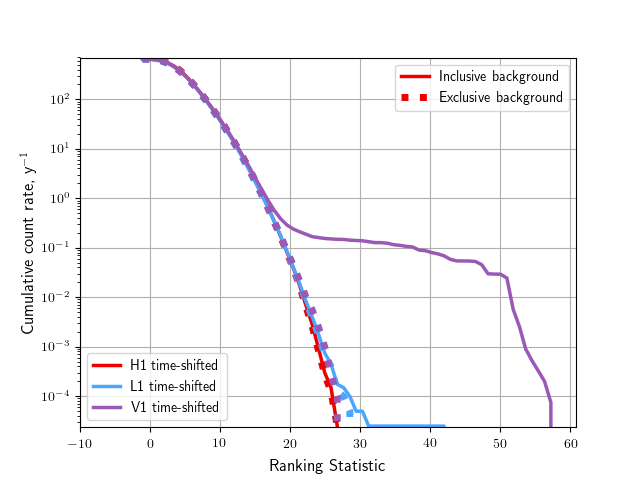}
    \caption{Histograms of ranking statistic for three-detector coincidences from the noise background, colored according to the detector that is time-shifted when forming the background estimates. The solid lines show the ranking statistic including all triggers; the dotted line shows background coincidences excluding triggers that form candidate coincidences with FAR $<$ 1 per 0.003 years in any detector combination.  We see that when Virgo is time-shifted (magenta) the inclusive background (solid line) has a significant number of loud background coincidences compared to the exclusive background (dotted line), this is not observed when shifting LIGO Hanford or LIGO Livingston.
    }
    \label{fig:shift_choice}
\end{figure}
Figure~\ref{fig:shift_choice} shows that when choosing the least sensitive detector (Virgo) to be time-shifted, many background coincidences are produced at large values of the ranking statistic. This is due to one or more high-SNR candidate HL coincidences in the fixed sub-network matching to noise triggers in Virgo when time-shifted, contaminating the background. For the data chosen, there is a candidate HL coincidence due to GW170809~\cite{Abbott_2019}.

This effect is not seen in the exclusive background, which has had all triggers which form candidate coincidences with false alarm rate below a specified threshold in any combination removed, so representing the background distribution under the assumption that all such candidate coincidences are due to signals. The effect is also not seen in the cases with LIGO Hanford and LIGO Livingston as the shifted detector as the triggers from candidate HL coincidences are being separated by the time-shift and candidate HV and LV coincidences generally have lower value of the ranking statistic due to the lower sensitivity of Virgo. This justifies our choice of including the least sensitive detector in the fixed sub-network. The choice between shifting either LIGO Hanford or LIGO Livingston data is therefore unimportant; we choose to shift LIGO Hanford because it is first alphabetically.

\section{Multi-detector ranking statistic}
\label{s:rankingstat}
In order to compare different coincidences in different detector combinations to one another, we develop here a new ranking statistic which reflects our degree of belief that a candidate coincidence is astrophysical in origin and which is consistent between detector combinations. This ranking statistic is based on the rates of noise events in each detector (similarly to~\cite{cannon2015likelihoodratio}), in addition to measures of detector network sensitivity and multi-detector signal consistency describing the signal event rate .

The Neyman-Pearson optimal detection statistic for triggered searches is given by the ratio of the signal and noise event rate densities~\cite{Biswas:2012tv}. Accurate description of these rates is therefore required.

A coincident event of unknown origin can be described by parameters $\vec{\kappa}$, including the trigger SNR $\rho$, signal-glitch discriminator(s) $\chi^2$ and template sensitivity $\sigma$ for each participating detector labelled by $a$; the template intrinsic parameters, $\vec{\theta}$, comprising binary component masses and spins; and the amplitude ratio $\mathfrak{A}_{ab}$, time difference $\delta t_{ab}$ and phase difference $\delta\phi_{ab}$ between each detector pair labelled by $a \neq b$,\footnote{For more than one pair of detectors, the amplitude ratios and time and phase differences are not all independent variables.}
\begin{equation}
 \vec{\kappa} = \left\{ \left[ \rho_a, \chi^2_a, \sigma_a \right]\!, \vec{\theta}\,, [\mathfrak{A}_{ab}, \delta t_{ab}, \delta \phi_{ab}] \right\}.
\label{eq:definekappa}
\end{equation}
A ranking statistic $\Lambda$ is a function of this set of parameters. The optimal statistic maximises the expected number of coincident events due to signals that are recovered above a statistic threshold $\Lambda^*$ corresponding to a given FAR. 

The false alarm rate as a function of the threshold $\Lambda^*$ is
\begin{equation} 
\mathrm{FAR}(\Lambda^*) = \int r_N(\vec{\kappa}) \Theta(\Lambda(\vec{\kappa}) - \Lambda^*)\,d^n\vec{\kappa},
\label{eq:FARLambda}
\end{equation}
where $r_N$ is the noise event rate density over $\vec{\kappa}$; the true alarm (signal) rate density is given by the equivalent calculation replacing $r_N$ with $r_S \equiv \mu_S \hat{r}_S(\vec{\kappa})$, where $\mu_S$ is an astrophysical coalescence rate per volume per time and $\hat{r}_S(\vec{\kappa})$ is a transfer function describing the recovered event distribution, which will depend on detector orientation and sensitivities. 

The optimal detection statistic is given by the ratio of event densities
\begin{equation} 
\Lambda_\mathrm{opt}(\vec{\kappa}) = \mu_S \frac{\hat{r}_S(\vec{\kappa})}{r_N(\vec{\kappa})},
\end{equation}
and the problem reduces to finding the form of the signal and noise event rate density distributions. The overall signal rate $\mu_S$ is assumed constant, thus does not need to be known.

We describe in sections~\ref{ss:rankstatnoise} and~\ref{ss:rankstatsignal} how the noise and signal event rate densities respectively are calculated. By using the ratio of signal to noise rate densities as our ranking statistic, we ensure that it is comparable across different detector combinations. Considering the dynamic range of expected rate densities spans many orders of magnitude, it is convenient to use the logarithm of the ratio of signal and noise rate densities,
\begin{equation}
R = \log(r_{s, i}) - \log(r_{n, i}).
\label{eqn:rankstatstart}
\end{equation}

\subsection{Noise Model}
\label{ss:rankstatnoise}
Here we describe the methods used in estimating the coincident noise rate density, which involves calculating the distribution of triggers with a given re-weighted SNR statistic in the individual detectors, and combining the single-detector trigger rates to find the rates of noise coincidences of all possible types.

\subsubsection{Single-detector trigger distributions}
\label{sssec:sngl_fitting}
We estimate the rate of triggers with a given re-weighted SNR for each detector by fitting the overall distribution of triggers to a decreasing exponential function.

In Gaussian noise we would be able to analytically fit the noise distributions, and if the noise distributions in all detectors were Gaussian, we would be able to combine the noise distribution of SNR for each detector $d$, $p(\rho_d | N)$ in the form $\rho_c \equiv \sum_d \rho_d^2$ for coincident triggers. However due to the presence of glitches and non-Gaussian behaviour in the data, we are unable to do this.

We use the chi-squared discriminant of~\cite{PhysRevD.71.062001} to ensure that candidate events have frequency evolution consistent with a binary merger signal. Although we cannot analytically predict the $\chi^2$ distribution for glitch triggers in each detector, we can describe the density of noise triggers via a combination of $\rho$ and $\chi^2$,
\begin{equation}
\hat{\rho} = 
\begin{cases}
 \frac{\rho}{[(1 + (\chi^2_r)^{p/2})/2]^{1/p}} & \mathrm{for\,} \chi^2_r > 1 \\
 \rho & \mathrm{for\,} \chi^2_r \leq 1
\end{cases},
\label{eq:newsnr}
\end{equation}
where the `index' $p$ is usually set to 6. 
This re-weighted SNR, which is approximately equal to $\rho$ for reduced $\chi^2$ values close to 1, can be used to describe the distribution of single-detector triggers as a simplification relative to directly modelling the density in the $\rho$--$\chi^2$ plane~\cite{Cannon:2012zt}. 
We also implicitly include the sine-Gaussian veto as described in~\cite{Nitz_2018a,Nitz_2019} in our re-weighted SNR in order to down-weight `blip' glitches in the data.

In each template, $\theta$, and each detector $d$, we model the distribution of the rate of triggers $r_{d}$ with respect to $\hat\rho$
as a falling exponential, above a threshold $\hat{\rho}_{th}$:
\begin{equation} 
r_{d}(\hat{\rho};\vec{\theta}; N) = \mu(\vec{\theta})p(\hat{\rho} | \vec{\theta}, N),
\end{equation}
where
\begin{equation}
p(\hat{\rho} | \vec{\theta}, N) =
\begin{cases}
\alpha ( \vec{\theta}) \exp \left[-\alpha(\vec{\theta}) (\hat{\rho} - \hat{\rho}_{th})\right] & \hat{\rho} > \hat{\rho}_{th} \\
0 & \hat{\rho} \leq \hat{\rho}_{th},
\end{cases}
\label{eq:noisemodel}
\end{equation}
given model parameters of $\vec{\theta}$, the number of triggers in the template above threshold, $\mu(\vec{\theta})$, and $\alpha(\vec{\theta})$, the exponential decay rate. Before performing the fit, we remove a fixed, small number of high-$\hat{\rho}$ triggers from each detector in order to mitigate possible bias due to loud signals.

To calculate the model parameter $\alpha$, we use a maximum likelihood (ML) fitting procedure. The log likelihood for obtaining a set of samples $\{\hat{\rho}\}$ is
\begin{equation}
\ln p\left( \{\hat{\rho}\}_d | \alpha, n_{tr} \right)
 = n_{tr} \ln \alpha - \alpha \sum_j^{n_{tr}} (\hat{\rho}_j - \hat{\rho}_{d, th}),
\end{equation}
where $j$ is the index for each trigger, and $\hat{\rho}_{d,th}$ indicates that the fit threshold value of $\hat{\rho}$ could be different for each detector; in practice we use the same fit threshold for all detectors. This likelihood is maximized in each detector by 
\begin{equation}
\alpha_\textrm{ML} = (\bar{\hat{\rho}} - \hat{\rho}_{th})^{-1},
\end{equation}
where $\bar{\hat{\rho}}$ indicates the mean value of $\hat{\rho}$, and the fractional variance of the fit parameter is approximately $1/\sqrt{n_{tr}}$.

The $\alpha_\textrm{ML}$ and $n_{tr}$ values are calculated for each template. There are often relatively few triggers in each individual template, so the variance can be large. 
We reduce variance by then taking a moving average of the fit parameters over templates which have similar parameters, under the assumption that `nearby' templates will have similar noise distributions.  Since $\alpha_{\rm ML}^{-1}$ is a linear function of the mean $\bar{\hat{\rho}}$ for a single template, we may take a mean of $\alpha_{\rm ML}^{-1}$ values (weighted by $n_{tr}$) over several templates and obtain an identical result to performing the ML fit directly over all triggers in those templates. We also smooth the count of triggers above threshold by taking the mean over nearby templates.

This smoothing over template parameter was initially performed over templates with similar duration \cite{Abbott_2019}, however even at constant template duration the variation of fit parameters over the effective spin $\chi_\textrm{eff}$ and mass ratio $\eta$ parameters is not insignificant: therefore a multi-dimensional smoothing of $\alpha$ and $\mu$ is performed as in~\cite{2-ogc}.

\subsubsection{Coincident noise event rate estimation}

The optimal ranking statistic includes the expected rate density of coincident noise events, which we calculate from single-detector noise trigger rate densities.
The rate of noise coincidences in a template $i$ can be estimated by multiplying together single-detector noise trigger rates\footnote{In this section we do not explicitly include the dependence on the fitting parameters $\alpha$ and $\mu$, instead returning to the general notation $r_{di}(\hat\rho_d)$.} $r_{di}(\hat\rho_d)$ for each detector $d$, and the size of the window of allowed coincidences $A_{N\{d\}}$, where $\{d\}$ is the set of detectors involved in the coincidence.

The rate of noise coincidences for a set of detectors $\{d\}$ is
\begin{equation}
r_{\{d\}i} = A_{N\{d\}}\prod_d r_{di}(\hat{\rho}_d),
\end{equation}
where $r_{di}$ are the rates in the individual detectors for each template as a function of the reweighted SNR in that detector $\hat\rho_d$ and $A_{N\{d\}}$ is an allowed time window for forming coincidences.  
The time window is formed by the limits on trigger time differences $\delta t_{ab}$ between detectors $a$ and $b$. We denote these limits as $\tau_{ab}$ and specify them as the GW travel time between detectors plus a small allowance for timing error (typically $\approx \pm 2\,\textrm{ms}$).

The allowed time window for a two-detector coincidence is simply $A_{N\{12\}} = 2 \tau_{12}$. The allowed window for forming a three-detector event $A_{N\{123\}}$ via coincidence tests applied to each pair of detectors is a product of the times $\tau_{12}$ and $\tau_{13}$, subtracting terms corresponding to disallowed regions with $|\delta t_{23}| > \tau_{23} / 2$:
\begin{equation}
A_{N123} = 2 \tau_{12}\tau_{13} +2 \tau_{12}\tau_{23} +2 \tau_{13}\tau_{23}  - \tau_{12}^2 - \tau_{13}^2 - \tau_{23}^2.
\label{eq:N123}
\end{equation}
The time difference window populated by signals will be different to this because of limitations on combinations of time differences in addition to those on individual two-detector differences. For example, two-detector time differences cannot all take maximal values from the same signal unless all detectors are in a straight line (which is not the case for detectors located on Earth's surface), which is shown in figure~\ref{fig:allowedtimes}. This has the effect of restricting the bounding window to be an ellipse with area~\cite{Bhawal:1995bn}
\begin{equation}
A_{S123} = \pi \tau_{12} \tau_{13} \sin \psi_{23},
\label{eq:S123}
\end{equation}
where $\psi_{23}$ is the angle between the lines-of-sight to detectors 2 and 3 as measured at detector 1. Comparison of this result with Eq.~\eqref{eq:N123} given the HLV detector network shows us that around a third of noise coincidences will fall outside of this signal-populated area, as seen in figure~\ref{fig:allowedtimes}.

Since the effective time windows for signal and noise coincidences are thus of similar size, we neglect differences between them and incorporate the rate density of noise triggers in the ranking statistic as
\begin{equation}
\log(r_{n, i}) =  \log A_{N\{d\}} + \sum_d \log r_{di}(\hat{\rho}_d).
\label{eqn:noiserate}
\end{equation}

\subsection{Signal Model}
\label{ss:rankstatsignal}
The true rate of signals depends on the overall rate of astrophysical mergers, the location and orientation of sources, and the distribution of intrinsic signal parameters (masses, spins, etc.)\cite{LIGOScientific:2018jsj}. Although we do not consider the distribution of intrinsic parameters over sources in this work, it has been considered in~\cite{PhysRevD.89.062002,Nitz_2017} and employed in~\cite{2-ogc}. We instead look at the parameters we can use to predict the rate of recovered signals as a function of this (assumed constant) astrophysical merger rate.

\subsubsection{Source Location}
\label{sss:PTA}
Astrophysical populations of sources are expected to be isotropically distributed over sky location and binary orientation. Similarly, for sources in the nearby universe such as LIGO and Virgo are currently detecting, the population is expected to be nearly uniform in volume. We can estimate how this population prior impacts the distribution of sources which are detectable by the gravitational-wave network, and furthermore how it affects the observed distribution of signal amplitudes, phases, and times of arrival in different detectors. For example, the probability of finding a signal is not uniformly distributed over the time and phase difference between detectors, and the two are correlated~\cite{Nitz_2017}. 
Figure~\ref{fig:allowedtimes} shows a hard boundary to the allowed signal time differences for illustration purposes, but in reality there would be timing error which would blur the edges, and the time differences within the bounding ellipse would not be uniform. For more discussion of the shape of the prior histograms, see~\cite{Nitz_2017}.
In contrast the times of `arrival' of noise triggers will be random and uniformly distributed; similarly differences in gravitational wave signal phase between the detectors will also be uniformly distributed for noise coincidences. 

The ratio of signal amplitudes between detectors would also have different distributions for noise vs.\ signal coincidences: the relative amplitudes of signals in different detectors will be dependent upon the source's sky position and polarization angle.
 
For a given set of extrinsic signal parameters $\vec\Omega$
comprising the relative amplitudes $\mathfrak{A}_{\{d\}}$, time delays $\delta t_{\{d\}}$ and phase differences between each site $\delta\phi_{\{d\}}$, the probability for that set of parameters to be generated by a signal $p(\vec{\Omega}|S)$ forms part of our ranking statistic for coincident events. To find the probability distribution, we perform a Monte Carlo calculation given the detector locations to produce histograms which act as look-up tables for the probability density $p(\vec{\Omega}|S)$. As an improvement over~\cite{Nitz_2017}, we have updated these signal prior histograms to support a three-detector network for use in triple coincidences. We divide this by the expected probability density given noise coincidences $p(\vec{\Omega}|N)$, which we expect to be uniform over the parameter space.

\subsubsection{Detector sensitivity}
\label{sss:sigma}
The sensitive distance of a detector, defined as the luminosity distance at which a standard compact binary source has a given expected SNR, affects the expected rate of signals that produce triggers in a given detector. This distance varies substantially between detectors and over time: we include this information in our ranking statistic via a term accounting for network sensitivity for a given coincidence type.

The instantaneous sensitive distance in a given detector, for sources matching a template labeled by $i$, is proportional to the quantity $\sigma$ defined in~\cite{PhysRevD.85.122006}.\footnote{Technically, $\sigma$ is the expected SNR for a face-on binary coalescence with a waveform perfectly matching a given template, located directly overhead from the detector at a luminosity distance of 1\,Mpc.} Then, network sensitivity for a given coincidence type is determined by the least sensitive detector via $\sigma_{{\rm min}, i}$.  Under the assumption of a homogeneous distribution of sources in volume, the expected rate of signals for a given coincidence type is therefore proportional to $\sigma_{{\rm min}, i}^3$. 

To normalize this measure of instantaneous sensitivity we compare to the rate corresponding to representative values of $\sigma_i$ over the analysis time. For the LIGO-Virgo network, we choose as a representative $\sigma$ value the median network sensitivity for HL coincidences $\overline{\sigma}_{HL, i}$.\footnote{I.e.\ the minimum over H and L of the median detector sensitivity over observing time. In general, we will use the most sensitive coincidence type as representative for this normalization.} 
Thus, the time-dependent rate of signals in a given coincidence type described by $\sigma_{{\rm min}, i}$ is proportional to 
\begin{equation}
 r_{s, i} \propto \frac{p(\vec{\Omega}|S)}{p(\vec{\Omega}|N)} \frac{ \sigma_{{\rm min}, i}^3}{\overline{\sigma}_{HL, i}^3},
\label{eqn:signalrate}
\end{equation}

leading to a term in the (logarithm of) the relative rate of signal vs.\ noise triggers 
\begin{equation}
 R_{\sigma,i} \equiv 3 \left( \log \sigma_{{\rm min}, i} - \log \overline{\sigma}_{HL, i} \right).
 \label{eqn:rsigma}
\end{equation}
\begin{figure}[tb]
    \centering
    \includegraphics[width=\columnwidth]{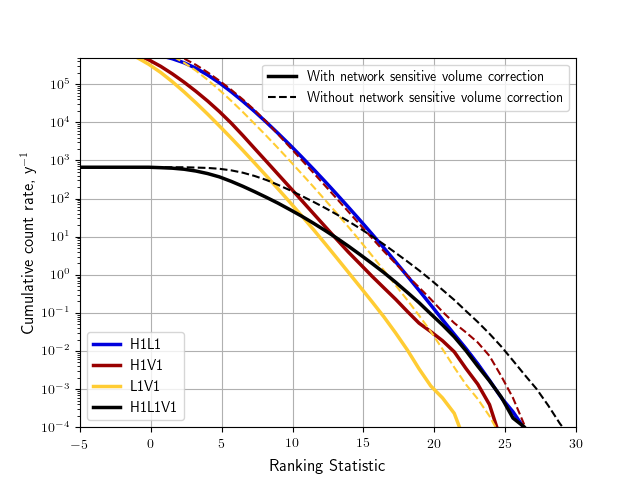}
    \caption{Histograms of ranking statistic for time-shifted coincidences from different combinations of detectors, colored by detector combination. The solid lines show the ranking statistic including the network sensitive volume term $R_\sigma$; dashed lines show the ranking statistic without this term.  We see that detector combinations containing Virgo (HV [crimson], LV [gold], HLV [black]) are penalized due to lower network sensitivity compared to HL [navy] which is not visibly affected by this term. Note also the much lower overall rate of triple (HLV) coincidences.
    }
    \label{fig:sigma_rankstatchange}
\end{figure}

Our statistic suppresses events in coincidence types where the least sensitive detector is significantly less sensitive than the others, as these are less likely to contain signals. During O2 the Virgo detector was much less sensitive than the LIGO detectors, thus as seen in figure~\ref{fig:sigma_rankstatchange} the distribution of background ranking statistics from HL coincidence is not significantly affected by the $R_\sigma$ term (median reduction of 0.00), whereas HV (2.62), LV (2.65) and HLV (2.68) statistic values are all heavily reduced. Note that coincidences in times of relatively poor sensitivity for H or L will also be penalized.

\subsection{Final ranking statistic}
\label{ss:rankstat}
Combining equation~\eqref{eqn:noiserate} for the noise rate density and~\eqref{eqn:signalrate} for the signal rate density into~\eqref{eqn:rankstatstart}, we then obtain our final ranking statistic
\begin{multline}
R = - \log A_{N\{d\}} - \sum_d \log r_{di}(\hat{\rho}_d) \\
+ \log p(\vec{\Omega}|S) - \log p(\vec{\Omega}|N) + R_{\sigma,i},
\label{eq:rankingstatistic}
\end{multline}
where $A_{N\{d\}}$ is the allowed time window for coincidence of equation~\eqref{eq:N123}, $r_{di}(\hat{\rho}_d)$ is the expected rate density of triggers in template $i$ and detector $d$ at re-weighted SNR $\hat{\rho}_d$, $p(\vec{\Omega}|S)$ is the probability of a signal having the extrinsic parameters $\vec\Omega$ given by the prior histograms, and $R_{\sigma,i}$ is (the log of) network sensitive volume for a given template and coincidence type, which is proportional to the expected rate of signals.

\section{Significance of multi-detector candidate events}
\label{s:significance}

The ranking statistic described above is designed to represent the relative probability of signal vs.\ noise origin for a coincident event, regardless of the detectors involved (see also~\cite{Biswas:2012tv,gstlal_3det,sachdev2019gstlal}). Thus we can compare statistic values across different coincidence types, considerably simplifying the task of producing a final list of candidates and calculating their significance.

To account for correlated events in different detector combinations produced by the same signal or noise transient, coincidences are clustered within a sliding window of ten seconds: the event with highest ranking statistic within a window, regardless of its detector combination, is kept and others are discarded. In order for this clustering operation not to damage search sensitivity, it is necessary for our statistic to correspond to the relative probability of signal vs.\ noise origin. 

This clustering over event types determines how false alarm rates are calculated in times where more than one available detector combination is active. To do this, we add together the estimated false alarm rates from all available detector combinations at the ranking statistic threshold of a candidate event. If our ranking statistic was not comparable across detector types, then we would not be able to calculate an equivalent false alarm rate in other detector combinations.

As an example, in the O2 data used in section~\ref{ss:a20comp}, if a coincident event is found from LIGO Livingston and Virgo with a ranking statistic of 10, this would have a FAR of approximately 1 per year. If LIGO Hanford data is not available at this time, then this would be the given FAR. But if LIGO Hanford is available, and did not participate in a more significant coincidence, then the false alarm rate for this ranking statistic in LV would be added to the FAR for HL coincidences (50 per year), HLV (0.8 per year) and HV (2 per year) at the same ranking statistic threshold to give an overall FAR  of around 54 per year. This combination method down-ranks triggers seen in less-sensitive detector combinations when more sensitive combinations are available.

This method affects the minimum FAR we can measure for coincidences in triple-detector time, which given the minimum measurable FAR in each coincidence type is
\begin{equation}
\textrm{FAR}_\textrm{min} = \frac{1}{t_{\textrm{bg,HLV}}} + \frac{1}{t_{\textrm{bg,HL}}} + \frac{1}{t_{\textrm{bg,HV}}} + \frac{1}{t_{\textrm{bg,LV}}},
\label{eq:FARmin}
\end{equation}
where $t_\textrm{bg}$ is the total time analyzed by time shifts in a given combination. Thus, for a three- (four-, five-) detector analysis, the FAR estimate in times where all detectors are observing is limited by approximately a factor four ($11$, $26$) relative to the minimal FAR in a comparable two-detector analysis. However as remarked above in Section \ref{ss:timeshifts}, the background time for combinations of three or more detectors can be extended at will by using more general multi-detector time shifts, thus only the double coincidence FAR estimates are truly limiting.

\section{Comparison of sensitivity}
\label{s:senscomp}

In order to compare the sensitivity of this analysis with the previous search, we estimate the total number of signals out of a notional (simulated) merger population that the two would detect at a given false alarm rate threshold.  We express the sensitivity of a search of a given data set as the product of volume of space and observing time, $VT$, under the assumption of mergers uniformly distributed in space and time: for a hypothetical signal population with local merger rate $\mu$, the expected number of signals the search would detect over the data set is $\mu VT$. Our figure of merit is therefore the ratio between these values for different analyses.

Since we do not know the true distributions of masses and spins of merging binary objects in the local Universe, we choose instead to calculate sensitivity for simple analytic distributions lying entirely within the parameter space of the search in order to simplify the interpretation of comparisons. We also neglect the effect of redshift on simulated signals (\textit{injections}), again to simplify the interpretation of sensitivity comparisons. Injections are performed for coalescences of binary black holes (BBH) and binary neutron stars (BNS); BBH injections use the waveform model SEOBNRv4~\cite{PhysRevD.95.044028}, and the BNS injections use the SpinTaylorT5 model~\cite{PhysRevD.84.084037}.  

An injection is considered to be detected if the highest ranked search event within one second of the injection merger time has an IFAR (inverse false alarm rate) value above a given threshold.  We then use importance sampling over many thousands of injections to calculate the sensitive volume for the target uniform-in-volume distribution~\cite{Usman_2016}.

We perform two comparisons to measure sensitivity improvements due to updates described here. The first is a comparison using colored Gaussian data at design sensitivity in section~\ref{ss:GaussianNoise}, and then an analysis of real O2 data in section~\ref{ss:a20comp}.

Our tests on Gaussian noise in section~\ref{ss:GaussianNoise} are designed to measure the improvement in sensitivity for future uses of the search given the presented improvement in the analysis: we analyse a fake three-detector network at design sensitivity and compare it to a network containing just the two LIGO detectors.

The tests on O2 data in section~\ref{ss:a20comp} are designed to show improvements in how we deal with non-Gaussian transient noise in the data, commonly referred to as `glitches'. These glitches cause spikes in SNR and can therefore increase the rate of coincident noise triggers. Real data is also non-stationary with significantly time-dependent sensitivity in each detector, which our statistic is designed to account for. 

\subsection{Colored Gaussian Data}
\label{ss:GaussianNoise}
By comparing injections in Gaussian data, we see the improvement in sensitivity by using three detectors rather than two in the PyCBC analysis. The sensitivity will be increased by both allowing coincidences where either of the LIGO detectors are not operating or does not produce a trigger, and by updates to ranking of signals seen within all three detectors during triple-detector coincident time. 

We use Gaussian noise which has been colored according to design sensitivity curves for advanced LIGO and Virgo~\cite{AdvLIGO2015, Acernese_2014}, and generate 107.3 total hours of coincident data over a five day period. We apply observational data segment definitions and data quality vetoes derived from a section of late O2 data in order to obtain realistic duty cycles of different two- and three-detector combinations and segment lengths for triggers generation. The coincident time available during this test is shown in table~\ref{t:coinctimes}.
\begin{table}[ht]
\centering
\begin{tabular}{|| c | c c ||} 
 \hline
  & \multicolumn{2}{c||}{Coincident time (hours)} \\
 Detector Combination & Fake Data & Real Data \\
 \hline
 HL & 86.5 & 129.6 \\ 
 HV & 89.7 & 108.1 \\
 LV & 95.3 & 120.2 \\
 HLV & 82.2 & 99.4 \\
 \hline
 Total coincident time & 107.3 & 159.2\\
 \hline
\end{tabular}
\caption{Table of times in the data used in the two tests shown in this paper for times which are coincident between LIGO Hanford (H), LIGO Livingston (L) and Virgo (V) respectively. Each of the HL, HV, LV times is inclusive of HLV time.}
\label{t:coinctimes}
\end{table}

The injections performed are separated into two bins: BNS injections with total mass between $2$ -- $5\,M_\odot$ and BBH injections with total mass between $5$ -- $100\,M_\odot$.  We note that since all templates have Gaussian noise event distributions in this data, we expect identical sensitivity to all signals that match the templates except for the fact that higher-mass systems produce higher amplitude GW signals at a given distance.  We choose to perform the importance sampling volume integral in a way that scales out this amplitude factor: thus our sensitivity estimate effectively weights every simulated signal equally at a given chirp distance\cite{PhysRevD.85.122006}, regardless of binary mass. We locate the simulated mergers uniformly on the sky and with uniform distribution in chirp distance between limits of 5\,Mpc and 600\,Mpc. 

For the BBH injections the logarithms of component masses are distributed uniformly between mass limits $2.5\,M_\odot$ and $50\,M_\odot$, while for BNS injections the component masses are uniformly distributed between 1 and $2.5\,M_\odot$. The component spins for BNS are distributed between 0 and 0.4, and in the BBH case, between 0 and 0.998. The BBH spins are strictly aligned with the orbital angular momentum, but the BNS spins are not.
\begin{figure}[htp]
    \centering
    {
    \hspace*{-0.5cm}
    \includegraphics[width=\columnwidth]{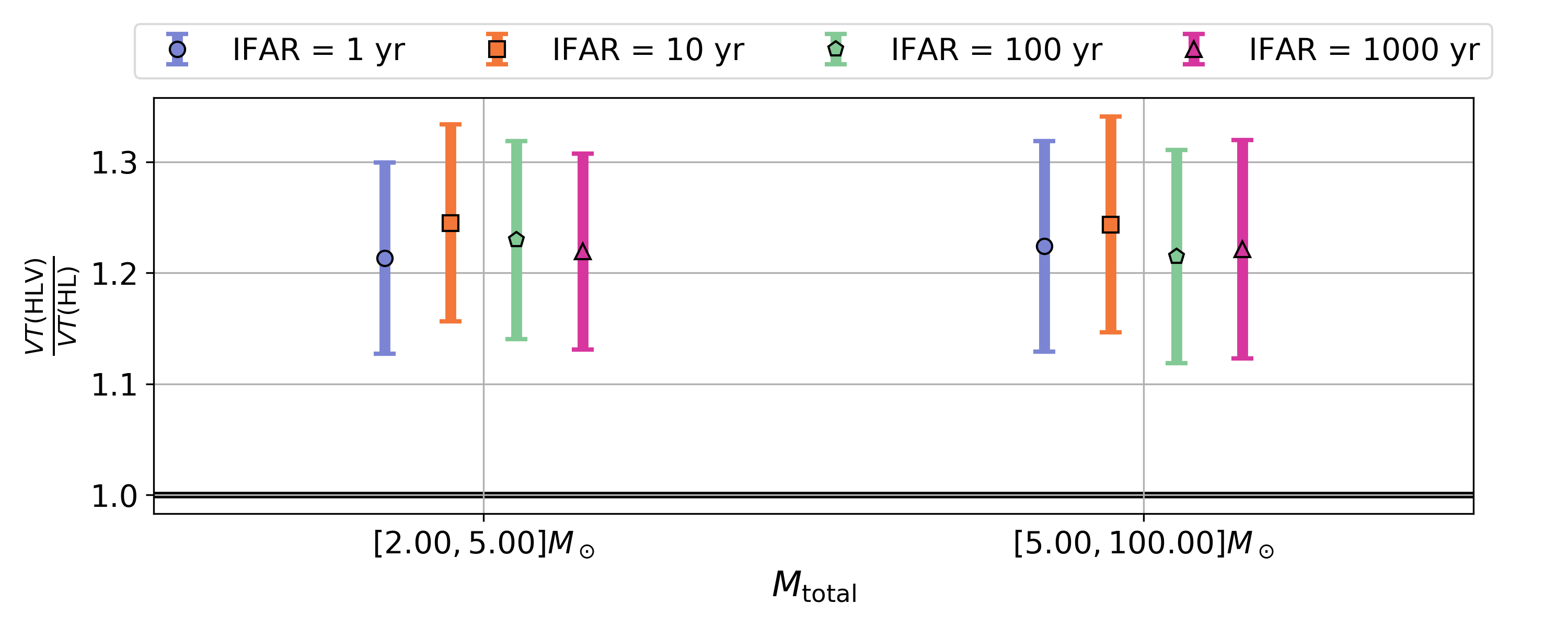}
    }
    \caption{Ratio of search sensitivities, $VT$, between the three-detector HLV analysis and two-detector HL analysis in colored Gaussian noise, plotted for four choices of IFAR threshold. Simulated signals used to estimate sensitivity have been split by total mass into BNS and BBH bins.
    }
    \label{fig:VTcomparisonFakeData}
\end{figure}

Figure~\ref{fig:VTcomparisonFakeData} shows an increase in $VT$ by a factor of $1.23 \pm 0.10$ for the three-detector analysis over the HL analysis for both source types, constant over all IFAR values considered.

A portion of this increase is due to HV or LV (two-detector) coincident time being available to the HLV analysis but not the HL analysis. We can approximate this factor via the total coincident analysis time divided by the HL time, weighted for the maximal volume sensitivity over available combinations at a given time. Using the sensitive range of an equal-mass binary black hole coalescence with component masses of $20\,M_\odot$, we find this factor to be approximately $1.12$. 

The remaining increase in $VT$ is then attributed to an increase of search sensitivity in three-detector time.
This is partly due to a subset of signals that generate HV or LV events in the three-detector search, which would not be seen in the two-detector HL search; also, to the much reduced rate of noise events for three-detector coincidence, implying that even relatively low SNR signals which generate a three-detector event will be more likely to be highly ranked and significant than in the two-detector case.

\begin{figure}[htp]
    \centering
    {
    \hspace*{-0.5cm}
    \includegraphics[width=\columnwidth]{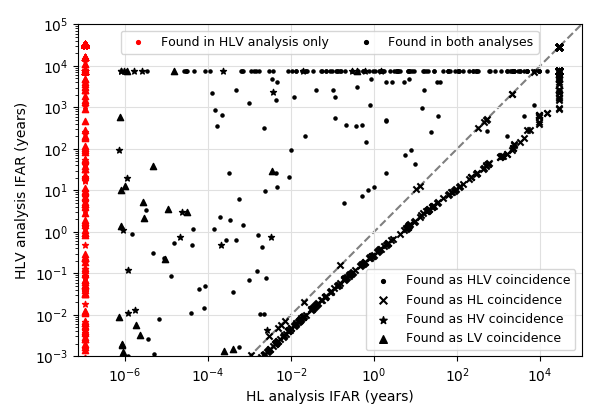}
    }
    \caption{
    Scatter plot of IFAR (inverse false alarm rate) for injections recovered in the HLV three-detector analysis vs the IFAR found in the two-detector HL analysis. The points are colored according to whether they were found in both the HLV and HL analyses (black) or if they were found in the HLV analysis only (red, plotted at a nominal HL analysis IFAR value). The different markers denote which detector combination the injection was recovered in during the HLV analysis. For 
    injections recovered in HL observing time there is no difference in the recovered significance (points lying on the gray line). No significant injections were seen in the two-detector HL analysis which were not seen in the HLV analysis.
    }
    \label{fig:fake_ifar_vs_ifar}
\end{figure}

We see in Figure~\ref{fig:fake_ifar_vs_ifar} that, in general, the significance increases for injections found in the HLV three-detector analysis when compared to the HL two-detector analysis significance.
We also see that the maximum obtainable IFAR for HLV events during triple-coincident time is less than that for other detector combinations, as no HLV coincident points have an IFAR above $10^4$\,years, but two-detector coincidences are all seen above this value, as described in equation~\ref{eq:FARmin}. There is a slight drop in significance for a population of HL triggers which fall in triple-coincident time for the same reason.

\subsection{Data from LIGO-Virgo Observing Run 2}
\label{ss:a20comp}
In the next test we use real GW detector data from the second observing run of Advanced LIGO and Virgo, O2, from between 2017-08-05 and 2017-08-13. This shows the response of the analysis to signals in the presence of noise artifacts including non-Gaussian transients and time-varying detector sensitivity. Again, we characterise the difference in sensitivity through the $VT$ ratio between two analyses: in this case we compare the HLV analysis with the HL analysis performed for the GWTC-1 catalog~\cite{Abbott_2019}. We see the amount of data used for this analysis in table~\ref{t:coinctimes}.

The injections used for this test are located uniformly on the sky and with uniform distribution in chirp distance between limits of 5\,Mpc and 300\,Mpc. This maximum distance is much less than that used in the fake data case, due to the difference in sensitive distance between the two example data sets. The BNS component masses are distributed uniformly between $1\,M_\odot$ and $3\,M_\odot$, and BBH are distributed uniformly in total mass, and then in primary mass for constant total mass, with component masses between $2\,M_\odot$ and $98\,M_\odot$ up to a maximum total mass of $100\,M_\odot$. 
The spins for injections in this analysis are distributed in the same way as the fake data injections described above.

We separate the injections used here into four bins between 2, 5, 16, 50 and 100 $M_\odot$ total mass. The motivation of this split is that different parts of the template bank are affected differently by non-Gaussian and non-stationary noise, thus we might expect injections recovered in various mass ranges to be differently affected by changes to the analysis.
\begin{figure}[htp]
    \centering
    {
    \hspace*{-0.5cm}
    \includegraphics[width=\columnwidth]{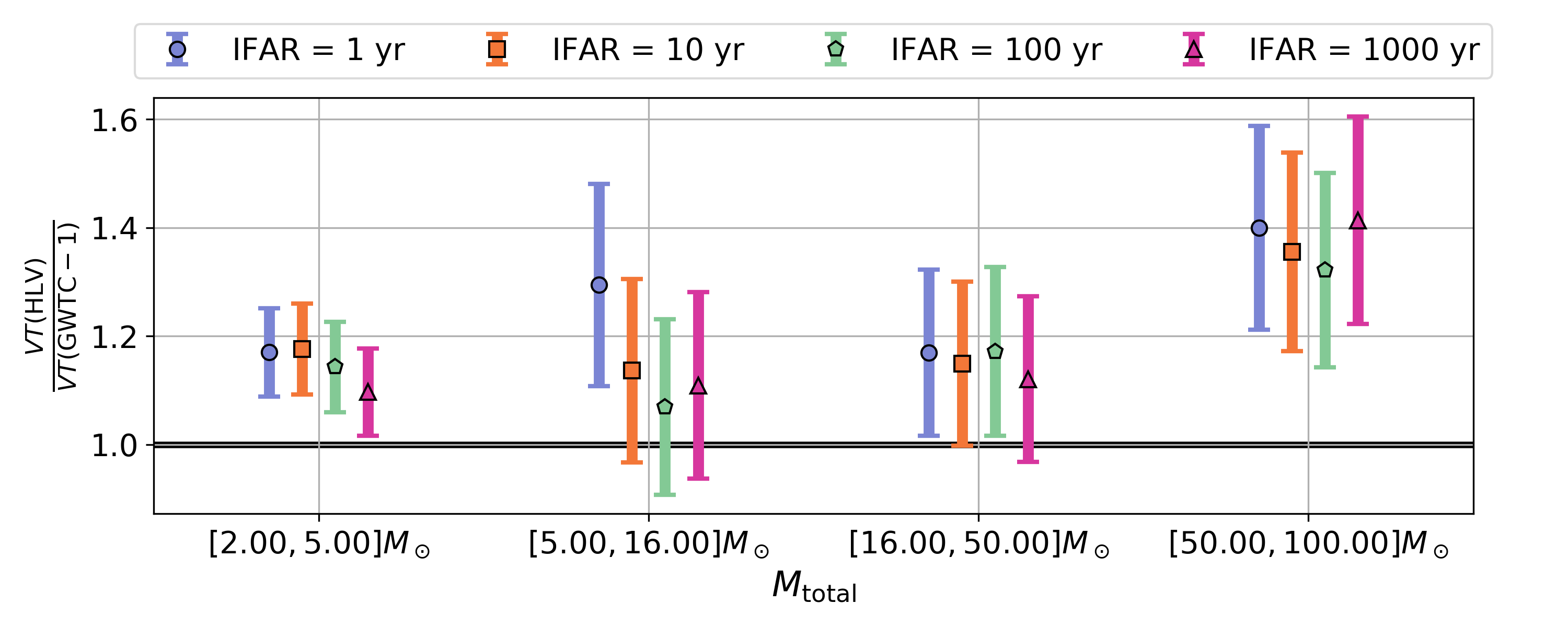}
    }
    \caption{Volume $\times$ time ($VT$) ratio comparing analyses of real data with the updated analysis against the PyCBC analysis as performed in~\cite{Abbott_2019}. There is a significant increase in $VT$, particularly for signals from heavy binary black holes, which may be due to changes to methods to differentiate noise artefacts from signals.}
    \label{fig:VTcomparisonReal}
\end{figure}

Figure~\ref{fig:VTcomparisonReal} shows an increase in sensitivity of between a factor of $1.14 \pm 0.08$ and $1.37\pm 0.18$, depending on the mass of the system, averaged over the different IFAR values. The strong dependence on the masses of the binary may be correlated to the presence of glitches which mimic the gravitational wave signature of very heavy binary black hole mergers~\cite{Cabero_2019}. By down-ranking coincidences which do not fit the time, phase and amplitude differences of section~\ref{sss:PTA}, we greatly reduce the ranking statistic of the glitch background in heavy BBH templates: thus, our injected signals will be seen with higher significance, and therefore to a greater distance for a given IFAR threshold.

As before, some of the increase in sensitivity is due to increased observation time, however in this case the expected increase in $VT$ is only a factor $\sim 1.01$, rather than the factor $1.12$ estimated in section~\ref{ss:GaussianNoise}, despite a similar relative increase in the analysis time. This is because, unlike the fake data case, here the Virgo detector has a sensitive range less than half that of the LIGO detectors, thus the expected sensitivity in times when only HV or LV are observing is much smaller than for HL times.

\section{Conclusions}
\label{s:conclusion}
We have presented changes to the PyCBC offline coincidence search related to analysing data from more than two detectors within the same analysis. These changes mean that we can now search over many detector combinations, and take certain characteristics of detector behaviour into account within the offline search. These improvements also mean that we can suppress noise coincidences more than before, and improve our prospects of finding signals within the data.

Tests on Gaussian data have shown that by using more detectors in the analysis for a LIGO-Virgo network at design sensitivity we increase the sensitive $VT$ by a factor of $1.23 \pm 0.10$. This is largely due to better ranking of signals seen in all detectors during triple-detector time due to suppression of noise, but also partly due to an increase in the duty factor of the network, even if the network has a slightly lower sensitivity. This second factor is not as significant if the additional detectors do not have relatively equal sensitivities, but with improvements to the Virgo detector ongoing and making its sensitivity more comparable to the LIGO detectors, we expect that this will become more significant in the near future.

We have also shown that the changes made to the analysis deal better with non-Gaussian noise realisations in the data due to better signal consistency checking and noise suppression. This means that in O2 data, where the Virgo detector was available but not significantly sensitive, we obtain an increase in $VT$ sensitivity by a factor of between $1.14 \pm 0.08$ and $1.37\pm 0.18$, depending on source properties. The greatest increase in sensitivity in this test was for signals from heavier black hole binaries, this may be due to re-weighting of templates which had previously been largely affected by specific types of detector glitches.

A full catalogue of the gravitational wave events identified in the O1 and O2 runs using the search described here is available in~\cite{2-ogc}.

\section{Further Work}
As future work, the coincidence tests and ranking method developed here could be applicable to increase the sensitivity of the the PyCBC low latency search, or be of interest to other CBC analysis pipelines looking to extend their searches to three- or more-detector analysis.

The new method for calculation of prior histograms can support an arbitrary number of detectors, however, memory requirements may become impractical for larger numbers of detectors. Alternate methods such as taking incoherent combinations of two/three detector prior histograms may provide accurate enough modelling to handle these cases and will be tested in the future.

To ensure that our event ranking is close to optimal, the modelling of noise trigger distributions as in section~\ref{sssec:sngl_fitting} may require updated fitting models. As significant multi-ifo events may involve triggers with lower SNR, the distribution of these triggers is affected by SNR thresholding, and is not easily fit by a simple function for all values of $\hat{\rho}$. Future work will investigate more accurate models for such distributions.

\section{Acknowledgements}
\label{s:acknow}
GSD and TD acknowledge support from the Maria de Maeztu Unit of Excellence MDM-2016-0692 and Xunta de Galicia. The PyCBC offline search software is built upon LALSuite~\cite{lalsuite}, numpy~\cite{numpy}, SciPy~\cite{Virtanen:2019joe} and Astropy~\cite{Price-Whelan:2018hus}.

The analyses in this paper used PyCBC version 1.16.2~\cite{alex_nitz_2020_3874649}. The authors are grateful for computational resources provided by the LIGO Laboratory and supported by National Science Foundation Grants PHY-0757058 and PHY-0823459.

\bibliography{references}
\end{document}